\documentclass[%
reprint,
superscriptaddress,
 amsmath,amssymb,
 aps,
 pra,
]{revtex4-2}

\usepackage[T1]{fontenc}
\usepackage[english]{babel}
\usepackage[colorlinks=true, citecolor=blue,urlcolor=blue,linkcolor=blue,filecolor=black]{hyperref}
\usepackage{physics}
\usepackage{graphicx}

\newcommand{\beq}{\begin{equation}}
\newcommand{\eeq}{\end{equation}}
\renewcommand{\emph}{\textit}
\renewcommand{\tr}{\text{Tr}}

\usepackage[utf8]{inputenc}
\usepackage{mathptmx}
\usepackage{mathtools}

\newcommand\coolover[2]{\mathrlap{\smash{\overbrace{\phantom{%
    \begin{matrix} #2 \end{matrix}}}^{\mbox{$#1$}}}}#2}

\newcommand\coolunder[2]{\mathrlap{\smash{\underbrace{\phantom{%
    \begin{matrix} #2 \end{matrix}}}_{\mbox{$#1$}}}}#2}

\begin{document}

\preprint{AIP/123-QED}

\title[]{
Generalized Time-bin Quantum Random Number Generator with Uncharacterized Devices
}

\author{Hamid Tebyanian}
\address{Department of Mathematics, University of York, Heslington, York, YO10 5DD, United Kingdom}

\author{Mujtaba Zahidy}%
\affiliation{ 
Centre of Excellence for Silicon Photonics for Optical Communications (SPOC), Department of Electrical and Photonics Engineering, Technical University of Denmark, Kgs. Lyngby, Denmark
}%

\author{Ronny M\"uller}%
\affiliation{ 
Centre of Excellence for Silicon Photonics for Optical Communications (SPOC), Department of Electrical and Photonics Engineering, Technical University of Denmark, Kgs. Lyngby, Denmark
}%

\author{S{\o}ren Forchhammer}%
\affiliation{ 
Centre of Excellence for Silicon Photonics for Optical Communications (SPOC), Department of Electrical and Photonics Engineering, Technical University of Denmark, Kgs. Lyngby, Denmark
}%

\author{Davide Bacco}%
\affiliation{ 
Department of Physics and Astronomy, University of Florence, Via Sansone 1, Sesto Fiorentino, 50019, Italy
}%

\author{Leif. K. Oxenl{\o}we}%
\affiliation{ 
Centre of Excellence for Silicon Photonics for Optical Communications (SPOC), Department of Electrical and Photonics Engineering, Technical University of Denmark, Kgs. Lyngby, Denmark
}%

\begin{abstract}
Random number generators (RNG) based on quantum mechanics are captivating due to their security and unpredictability compared to conventional generators, such as pseudo-random number generators and hardware-random number generators.  
This work analyzes evolutions in the extractable amount of randomness with increasing the Hilbert space dimension, state preparation subspace, or measurement subspace in a class of semi-device-independent quantum-RNG, where bounding the states' overlap is the core assumption, built on the prepare-and-measure scheme. We further discuss the effect of these factors on the complexity and draw a conclusion on the optimal scenario.
We investigate the generic case of time-bin encoding scheme, 
define various input (state preparation) and outcome (measurement) subspaces, and discuss the optimal scenarios to obtain maximum entropy. Several input designs were experimentally tested and analyzed for their conceivable outcome arrangements. We evaluated their performance by considering the device's imperfections, particularly the after-pulsing effect and dark counts of the detectors. 
Finally, we demonstrate that this approach can boost the system entropy, resulting in more extractable randomness.
\end{abstract}

\maketitle

\section{Introduction}
\label{Sec::Intro}
Randomness is indispensable for simulation, gambling, and numerous cryptographic applications, e.g., quantum key distribution (QKD) \cite{QKD,Zahidy_2021_OAM}, where the protocol's security is guaranteed by random selections of the encoding and measurement bases \cite{Acin2016}. Traditional randomness generators rely on deterministic processes, which are, in principle, predictable. However, unlike the deterministic evolution of classical systems, quantum mechanics grants the ability to generate genuine randomness based on the quantum measurement outcome that is entirely unpredictable \cite{rev_QRNG,julio_RNG_Tests}. A random number generator (RNG), in general, should deliver unpredictable and secure random numbers by exploiting effective instruments aiming to make it performant, high rate, and commercially affordable. Quantum RNG (QRNG) can be an outstanding choice in satisfying the needs for security, practicality, and affordability; nevertheless, any imperfection in the physical realization may cause information leakage which an eavesdropper could use to predict the QRNG's outcome \cite{Stipcevic12,Herrero-Collantes2017}.

Nowadays, QRNGs are commercially available, symbolizing one of the most successful developments of quantum technologies. 
In Device-dependent (DD) QRNGs, the user must trust the device's performance. This type of QRNG requires a detailed understanding of the functioning of the in-use devices to constrain the output's randomness \cite{zahidy2021quantum,DD_IDQ,DD_marco,QRNG_gen-new}.
Although DD QRNGs randomness is guaranteed by quantum theory, any gap between theoretical and real-world implementation, such as experimental errors, device imperfections, or dishonest producers, may enable an adversary to predict the QRNG's outcomes and thus endanger the system's security \cite{DI_roger,colbeck2011quantum,Colbeck2011,DI_me,rutvij_DI}. At the same time, in device-independent (DI) protocols, one can certify randomness without relying on assumptions about the device's performance. These protocols utilize the non-local property of quantum theory to guarantee the output's randomness. DI QRNGs are, therefore, highly secure, and thus no assumptions on the eavesdropper are made. Implementing DI QRNGs, nevertheless, can be demanding as it involves conducting a loophole-free Bell test, which is a challenging experimental task with a typically low generation rate \cite{DI_new}.

Contrary to DD and DI QRNGs, semi-device independent QRNGs are based on protocols that allow for high-rate generation, acceptable security, and simplicity in implementation \cite{semi-DI,MDI_new,semi_DI_new,SDI_prx}. In this class, the performance is boosted by taking a few assumptions on the working principle of the experimental apparatus, e.g., trusting the measurement \cite{sdi_neww,SDI_Avesani2022} or the preparation device \cite{MDI_new,MDIII} or weaker hypothesis like bounding the energy or the overlap \cite{Tebyanian_2021,Brask2017} of the generated states, while guaranteeing the security by accounting for all possible attack attempts within our assumptions \cite{Ma2016}.

\begin{figure*}[ht]
    \includegraphics[width=\linewidth]{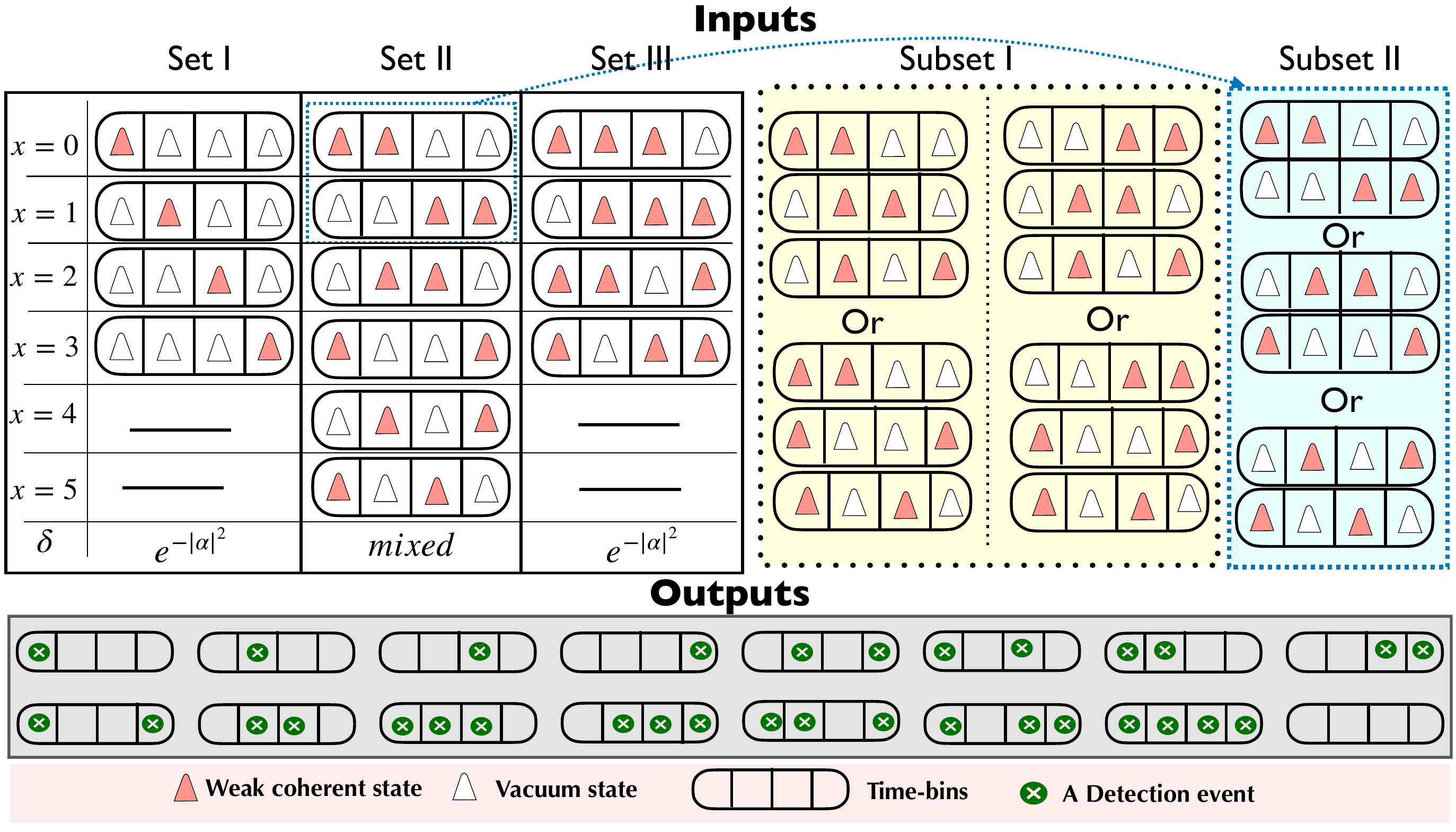}
    \caption{Possible input-output configurations with four time-bin case. Inputs: Sets I, II, and III  show input configurations where one, two, and three weak coherent states are positioned in time-bins, respectively. Subsets I and II are subsections of set II where the overlap is mixed. Outputs: 16 possible outcome configurations for four time-bin case, where some are theoretically impossible, e.g., obtaining four detection events, while real-world errors such as the detector's dark counts make it probable.}
    \label{Fig::PermutStates}
\end{figure*}

This work studies a class of semi-DI QRNGs founded on the basis of restraining the states' overlap by employing a time-bin encoding scheme and single-photon detection. The overlap bound guarantees that the prepared states are non-orthogonal and hence, no measurement can perfectly distinguish them \cite{VanHimbeeck2017semidevice, Brask2017}. While the inability of predicting the outcome of measurement by the user is the source of randomness, the indistinguishability of the state is the source of security, from the perspective of the measurement apparatus. The entropy and extractable randomness are optimized, and compared, with the help of semi-definite programming (SDP). We discuss the improvement in entropy and randomness generation rate with increasing the number of time-bin or input states.

The main contribution of this work is to investigate the impact of increasing or adjusting the number of time bins on the extractable amount of randomness and the system's generation rate with the security assumption. We found an upper bound on the number of input-output for a general number of time bins and showed that the system's entropy improves with a increasing number of time bins. We also discuss the experimental challenges from both state preparation and measurement points of view. Similarly, we demonstrate that the generation rate increases by optimally dispersing the weak coherent state (WCS) in time-bin configurations, which can significantly enhance this approach's performance for practical applications.

\section{Protocol}
\label{SEC::Protocol}
The QRNG protocol introduced here is based on the prepare-and-measure scenario, where the prepared states' overlap is bounded while no other assumptions are required on the rest of the setup \cite{tpsk_QPSK,Tebyanian_2021,avesani2020}. 

\subsection{Preparation and Measurement Stages}
\label{SubSEC::PrepMeas}
Quantum mechanics does not allow any measurement to distinguish non-orthogonal states perfectly \cite{Barnett:09}. This feature can be used to generate random numbers without trusting the measurement apparatus. Here, we address a general case of non-orthogonal states in a time-bin encoding with $n$ bins and $m$ distributed weak coherent pulses $| \alpha \rangle$. The states $\ket{\psi_i}$, 
\beq
\ket{\psi_i} = |0\rangle^{n-m}  | \alpha \rangle^{\otimes m} = |0\rangle \otimes | \alpha \rangle \otimes ... \otimes |\alpha\rangle\otimes |0\rangle,
\eeq
are formed by permuting the $m$ WCSs in the $n$ bins where the rest are filled with vacuum states (VS). The states $\ket{\psi_i}$  are required to respect an overlap condition that satisfies the protocol's assumption:
\beq
|\bra{\psi_i}\ket{\psi_j}| \geq \delta,\quad \forall \; i\neq j,
\label{EQ::OverlapCond}
\eeq
where $\delta$ is the overlap bound. The non-zero overlap guarantees the inability to distinguish the states by performing any measurement, hence, allowing to generate secure randomness from the ambiguity therein \cite{Barnett:09}. A simple illustration of state formation in time-bin encoding can be found in \cite{Tebyanian_2021}.

In this scenario, the general case is defined by allowing the number of time-bins $n$ to increase without any limits as well as the number of WCSs $m$, where $1 \leq m < n$.
We denote a \textit{configuration} of $n$ time-bins and $m$ WCSs with $(n,m)$-configuration. The number of states in a $(n,m)$-configuration is given by the binomial coefficient, $C_n^m=n!/(m!(n-m)!)$, formed by all possible combinations of placing $m$ WCSs in $n$ time-bins. However, not all groups of states in a configuration respect the overlap bound, Eq. (\ref{EQ::OverlapCond}). A careful examination of combinations shows that in an $(n,m)$-configuration, there are subsets of states with specific overlaps. Each subset is then divided into groups of states that are equivalent w.r.t. the overlap value. Fig. (\ref{Fig::PermutStates}) shows the $(4,2)$-configuration and its subsets with different overlap values.
To be noted that while the four groups of subset I are not closed w.r.t. each other, adding any elements of another group to any of them violates the overlap bound.

It is easy to show that the number of subsets is equal to 
\[
  \begin{cases}
    m    & \quad \text{if } \quad 2m-n \leq 0 \\
    n-m  & \quad \text{if } \quad 2m-n > 0.
  \end{cases}
\]
Consequently, a $(n,m)$-configuration can have a total overlap value of the form
\begin{equation}
\begin{split}
|\bra{\psi_i}\ket{\psi_j}| & = \langle 0 | 0 \rangle^{n-2m+s} \langle 0 | \alpha \rangle^{2(m-s)} \langle \alpha | \alpha \rangle^s \\
& = \langle 0 | \alpha \rangle^{2(m-s)},
\end{split}
\label{over}
\end{equation}
where $s$ is the number of coinciding $\langle \alpha | \alpha \rangle$ WCSs. We denote an $(n,m)$-configuration with $s$ coinciding WCSs as $n_{m,s}$ with $n > m \geq s$.

In the following, we will only consider the case of equality in Eq. \eqref{EQ::OverlapCond}. We denote with $\mathcal{B}(n,m,s)$ the maximum number of states in any subset $\mathcal{S}$ of the $(n,m)$-configuration such that all elements in $\mathcal{S}$ have the same value of $s$ pairwise, with $s$ defined as in Eq. \eqref{over}. It is of relevance to know  $\mathcal{B}$ for any configuration as it defines the number of inputs and possible outputs in our prepare-and-measure QRNG protocol. This question is closely related to \textit{constant weight binary codes}. To see this, we can identify bins that contain a WCS with `1' and bins that contain the vacuum state with `0', such that we identify each state in a $(n,m)$-configuration with a binary vector of length $n$ and weight $m$. Each subset $\mathcal{S}$ can then be directly identified with a code of length $n$, Hamming distance $d$, and weight $m$, where Hamming distance and $s$ are related as $d=2(m-s)$. Eq. \eqref{over} can then be written as $|\bra{\psi_i}\ket{\psi_j}| = \langle 0 | \alpha \rangle^d$.  In the context of constant weight binary codes, there exists the well-known but open question of determining the maximum number of codewords $\mathcal{A}(n,m,d_{\text{min}})$, where $d_{\text{min}}$ refers to the minimum distance of the code. $\mathcal{B}(n,m,s)$ can be upper-bounded by $\mathcal{A}(n,m,2(m-s))$ which in turn can be upper-bounded by different theoretical bounds \cite{1057714, upp, schri}. Lower bounds to $\mathcal{A}$, typically by explicit construction \cite{59932, montemanni}, cannot be applied to $\mathcal{B}$ as the codes can contain state-pairs with $d>d_{\text{min}}$ which translates to a violation of Eq. \eqref{EQ::OverlapCond} since $\delta = \langle 0|\alpha \rangle^d$. Increasing $d$ reduces the overlap value and therefore reduces the ambiguity in their measurement. Instead, we show here an explicit lower bound $C$ by simple construction: For $2m-n\leq0$, all codewords share $s$ `1's at the same positions. Distribute the remaining $m-s$ ones in the remaining $n-s$ slots so that there is no coinciding ones, and fill the $R=n- \lfloor \frac{n-s}{m-s} \rfloor (m-s)-s$ leftover columns with zeros.\\
\vspace{0.2cm}
\[
\mathcal{B}=
\begin{bmatrix}
1   & ... & 1   &    \coolover{n-s}{       1    &   ...   &  1     & 0   & ... & 0      & 0 & ... & 0 & 0 &...&0}\\
... & ... & 1   &   0    &   ...   &  0     & 1   & ... & 1      & 0 & ... & 0&0&...&0 \\
... & ... & ... &  ...   &   ...   &  ...   & ... & ... & ...    & 0 & ... & 0 &0&...&0\\
\coolunder{s}{1&...&1} & \coolunder{m-s}{0&...&0}  & \coolunder{m-s}{0&...&0} & \coolunder{m-s}{1&...&1} & \coolunder{R}{0&...&0}\\
\end{bmatrix}
\]
\vspace{0.6cm}

This results in $C = \lfloor \frac{n-s}{m-s} \rfloor$ different states. If instead $2m-n>0$, all codewords share $n+s-2m$ zero positions and the remaining $2m-s$ slots are divided into sections with $m-s$ zeros. $\mathcal{B}$ can therefore be lower-bounded by
\begin{equation}
    \mathcal{B}(n,m,s) \geq C(n,m,s) = \begin{cases}
        1  &  \text{if} \quad n+s-2m < 0\\
        \lfloor \frac{n-s}{m-s} \rfloor  & \text{if} \quad 2m \leq n\\
        \lfloor \frac{2m-s}{m-s} \rfloor & \text{if} \quad 2m > n
    \end{cases}
\end{equation}

In the absence of noise or errors, the number of all possible outcomes, \textit{B}, follows from the click or no-click event when a state is sent. For an $n_{m,s}$-configuration, the number of distinct outcomes is obtained as 
\begin{equation}
    B = C (2^m - 1)- 2^{m-s} + 1.
    \label{EQ::NumOutcomes}
\end{equation}

In the no-frills case, only one WCS is placed, $m=1$, in each time-bin regardless of the number of bins, see Fig. \ref{Fig::PermutStates} (set I). There are always $B=n+1$ possible outcomes in this case — one for each input plus one for the no-click (indeterminate) event, which occurs randomly, suggesting that the entropy should be minimal. Fig. \ref{Fig::PermutStates} (Set II and III) shows the cases with $m=2$ and $m=3$, respectively. Note that the case with $m=2$ WCSs has two subsets with 1 and 2 coinciding WCSs with 4 equivalent groups for $m=2$ and 3 for $m=3$. In the ideal situation, the number of outcomes follows Eq. (\ref{EQ::NumOutcomes}). However, in a real implementation, due to noise, dark counts, or after-pulsing, all $B=2^n$ outcomes, shown in Fig. \ref{Fig::PermutStates} for $n=4$ - Outputs, are probable although with negligible probability. These errors and imperfections are viewed as classical side-information serving the adversary to predict the measurement outcome. All sorts of probable classical side-information and correlations (between preparation and measurement sides) are considered in the security estimation. The user can monitor these correlations and stop the protocol in case of observing considerable noise.

\subsection{Security estimation}
\label{subsec::SecEst}
Despite the fact that the generation of random numbers in a QRNG is based on the intrinsic probabilistic nature of quantum mechanics, the raw data outcome is a mixture of the sequences generated from deterministic classical sources and quantum processes. Therefore, it is essential to estimate the amount of extractable randomness in a defined protocol and later use it to exclude the classical contribution. The quantity min-entropy ($H_{\min}$) measures the maximum extractable randomness provided that an adversary can optimally guess the generator's outcome knowing the working principle of the devices. To account for any side information, we used conditional min-entropy and considered only classical side-information. Throughout this work, we assumed a trusty source with no quantum correlation to the outside world.

The conditional min-entropy on the variable $b$ conditioned on classical side-information $E$ reads \cite{Tomamichel2011}
\begin{equation}
    H_{\min}(b|E) = -\log_2 P_{\text{guess}}(b|E),
    \label{EQ::H_min}
\end{equation}
where $P_{\text{guess}}$ is the maximum probability that an adversary can guess the measurement outcome with a complete understanding of the devices' working principle and classical noises. In a semi-DI framework, the guessing probability should be maximized over all possible preparation and measurement strategies. $P_{\text{guess}}$ reads:
\beq
P_{\text{guess}}= \mathop {\max}\limits_{p(x),\psi_x,M_{b}^{{\varsigma}}}\bigg\{\sum\limits_{x = 0}^{I-1} p(x) \sum\limits_{{\varsigma} } \mathop{\max}\limits_{b}\big[\bra{\psi_x} M_{b}^{{\varsigma}} \ket{\psi_x}\big]\bigg\},
\label{P_gg}
\eeq
where $p(x)$ is the probability of transmitting input $x$, $M_{b}^{{\varsigma}}=P(\varsigma)\Pi_{b}^{{\varsigma}}$ are weighted measurement strategies over all positive operator valued measurements (POVM), and $\varsigma$, known by the adversary, represents the classical correlations between the measurement devices and environment (e.g., adversary). Each POVM $\Pi_{b}^{{\varsigma}}$, labeled by $\varsigma$, can be implemented with probability $P(\varsigma)$. $I$ and $B$ are the numbers of inputs and outcomes, respectively.
As shown in \cite{Bancal_2014}, the maximizations in Eq. (\ref{P_gg}) can be grouped as they occur for the same value of $b$ at given $x$, this would significantly ease up the optimization process. Therefore the total number of possible measurement strategies for given input would be $B^I$, thus $\varsigma \in \{\varsigma_0,\dots, \varsigma_{I-1}\}$, where  $\varsigma_s \in \{0,\dots.,B-1\}$. Following the same approach presented in \cite{Tebyanian_2021,Brask2017,d_out}, $P_{\text{guess}}$ for the balanced input case, $p(x)=1/I$, can be written as:
\beq
P_{\text{guess}} = \frac{1}{I}\mathop{\max }\limits_{\{M _b^{\varsigma},\hat\rho_x\}} \sum\limits_{x = 0}^{I-1}{\sum\limits_{\varsigma} \Tr[\hat\rho_x M _{\varsigma_x}^{\varsigma} ]},
\label{eq:P_g3}
\eeq
where $\hat\rho_x=\ket{\psi_x}\bra{\psi_x}$, and $\Tr[\hat\rho_x M _{b}^{\varsigma} ]=p(b|x)$ is the conditional probability of obtaining outcome $b$ given input $x$. 
Eq.(\ref{eq:P_g3}) suggests that $P_{\text{guess}}$ depends on the state's overlap rather than input state $\hat\rho_x$. Besides, the optimization problem in Eq.(\ref{eq:P_g3}) can be bounded to a $I$-dimensional Hilbert space; for more detail, see \cite{Brask2017,Tebyanian_2021,Bancal_2014}.

The optimization problem in $P_{\text{guess}}$ can be efficiently solved by casting it into semi-definite programming (SDP), which is a numerical tool for solving complex optimization problems.

Following the same argument presented in \cite{Tebyanian_2021,d_out,Bancal_2014,Brask2017}, we can show that for the protocol under study, strong duality holds which means both the primal and dual forms of the SDP exist. By feeding the SDP with the experimental conditional probabilities $P(b|x)$ and defining the overlap bound, the SDP can numerically optimize $P_{\text{guess}}$. Afterward, the conditional min-entropy, Eq. (\ref{EQ::H_min}), can be calculated.

\begin{figure}[h]
    \centering
    \includegraphics[width=\linewidth]{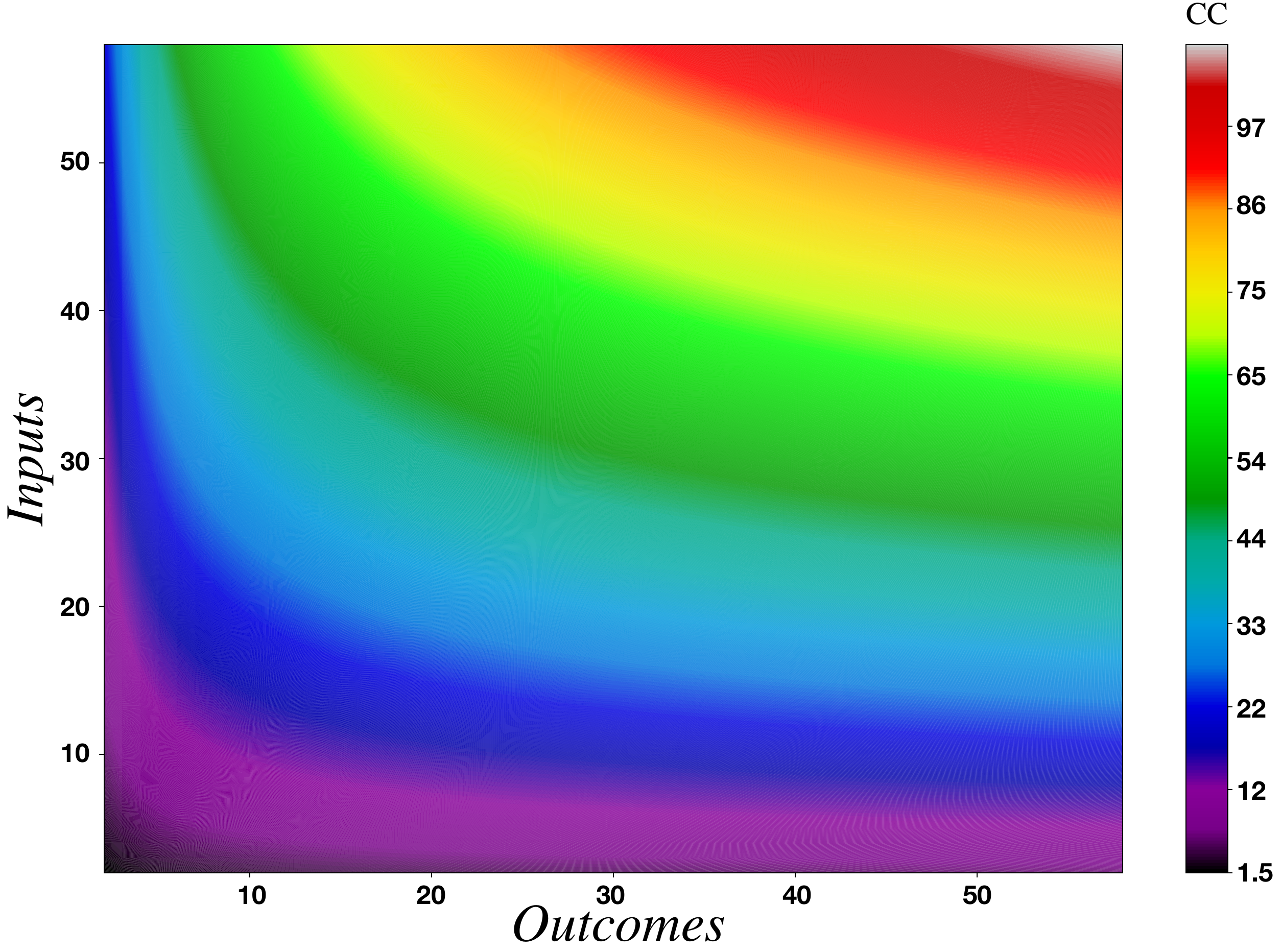}
    \caption{Computational cost (CC), colour bar, as a function of the number of inputs and outputs. Note that the CC is plotted on a logarithmic scale, expressing that CC increases exponentially with the number of IO.}
    \label{Fig::com}
\end{figure}

It should be noted that the security estimation is applicable for multiple input-output (IO) cases. The number of inputs can vary from 2 to the number of available states in an equivalence group in a $n_{m,s}$-configuration. For example, one can choose to send only 2 out of 4 states in set I in figure (\ref{Fig::PermutStates}). The computational cost (CC) is associated with the number of IO in the system and can affect the system's overall generation rate. This is due to an increment in the time it takes to execute the SDP, which in turn leads to a decrease in the system's overall efficiency. Thus, it is important to be mindful of the impact of increased computational complexity when considering adding more IO to the system. Fig. (\ref{Fig::com}) shows the CC as a function of the number of IO obtained on a personal computer.

Given a specific input, an outcome probability is a function of mean photon number per pulse $\mu$, detector efficiency $\eta_{\text{det}}$, noise in the form of background light, dark count, and after-pulsing. An approach to reduce the complexity of SDP is to group the outcomes, from an adversary point of view. This will drastically reduce the complexity of SDP.

It can be explained in a $n_{1,0}$-configuration where, in the absence of noise, there are $n+1$ different outcomes. The common outcome is the no-click one, and the others are 1-click due to the  WCS. In this case, a new variable ($E \in \{0,1\}$) can be assigned to the outcomes in which $E=0$ corresponds to the no-click event, all '0', while $E$ is 1 for $b\in\{\underbrace{100\dotsb 0}_\text{n}, 010\dotsm0, \dotsm, 0\dotsm01\}$.

\beq
\begin{aligned}
& P_{\text{guess}}= \mathop {\max}\limits_{p(x),\rho_x}\bigg\{\sum\limits_{x = 0}^{2} p(x) \\
& \sum\limits_{\varsigma_0,\varsigma_1,\varsigma_2 = 0}^{1} \mathop {\max} \big\{\Tr[\hat\rho_x M_{E=0}^{\varsigma_0,\varsigma_1,\varsigma_2 }],1-\Tr[\hat\rho_x M_{E=0}^{\varsigma_0,\varsigma_1,\varsigma_2 }]\big\}\bigg\}
\end{aligned}
\label{MHZ}
\eeq
For configurations with more WCSs more variables (corresponding to E) should be specified as there would be more indeterminate events.

The many-outcome approach is a computationally simplified, effective, and efficient method of increasing entropy without significantly increasing CC. 
This is a result of comparing the computational cost with increasing the number of inputs versus the number of outcomes which shows that the former increases faster, see Fig. (\ref{Fig::com}).
Hence, in an $n_{m,s}$ configuration, an efficient strategy is to keep the number of inputs fixed and low and increase the number of outcomes.

\begin{figure}[h]
\centering
\includegraphics[width=\linewidth]{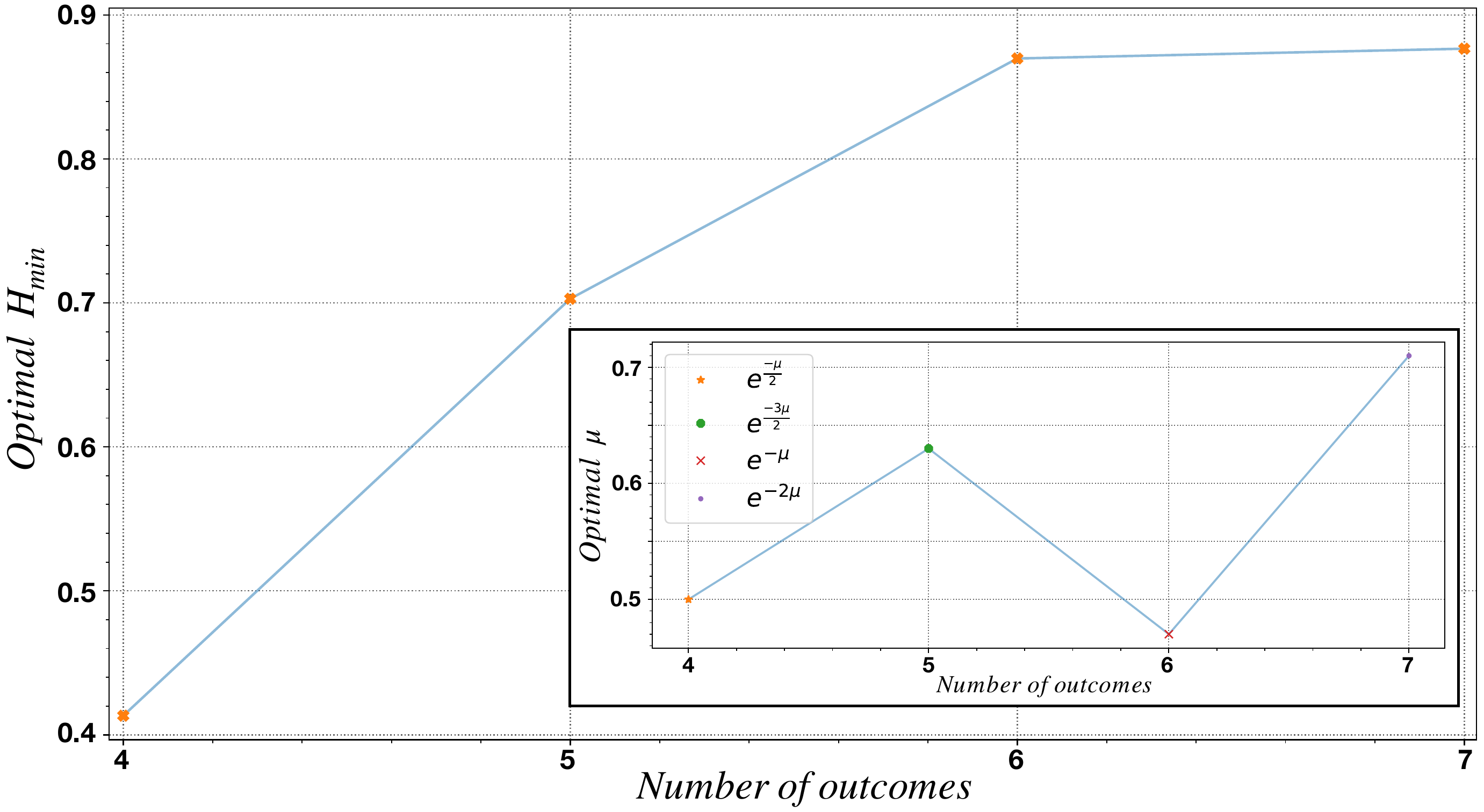}
\caption{Optimal conditional min-entropy as a function of the number of outcomes with binary input. \textit{Inset:} Optimal mean-photon number ($\mu$), i.e., the $\mu$ which delivers the maximum entropy, as a function of the number of outcomes for states with different overlaps.}
\label{Fig::optii}
\end{figure}

\begin{table}[htb]
    \centering
    \includegraphics[width=\linewidth]{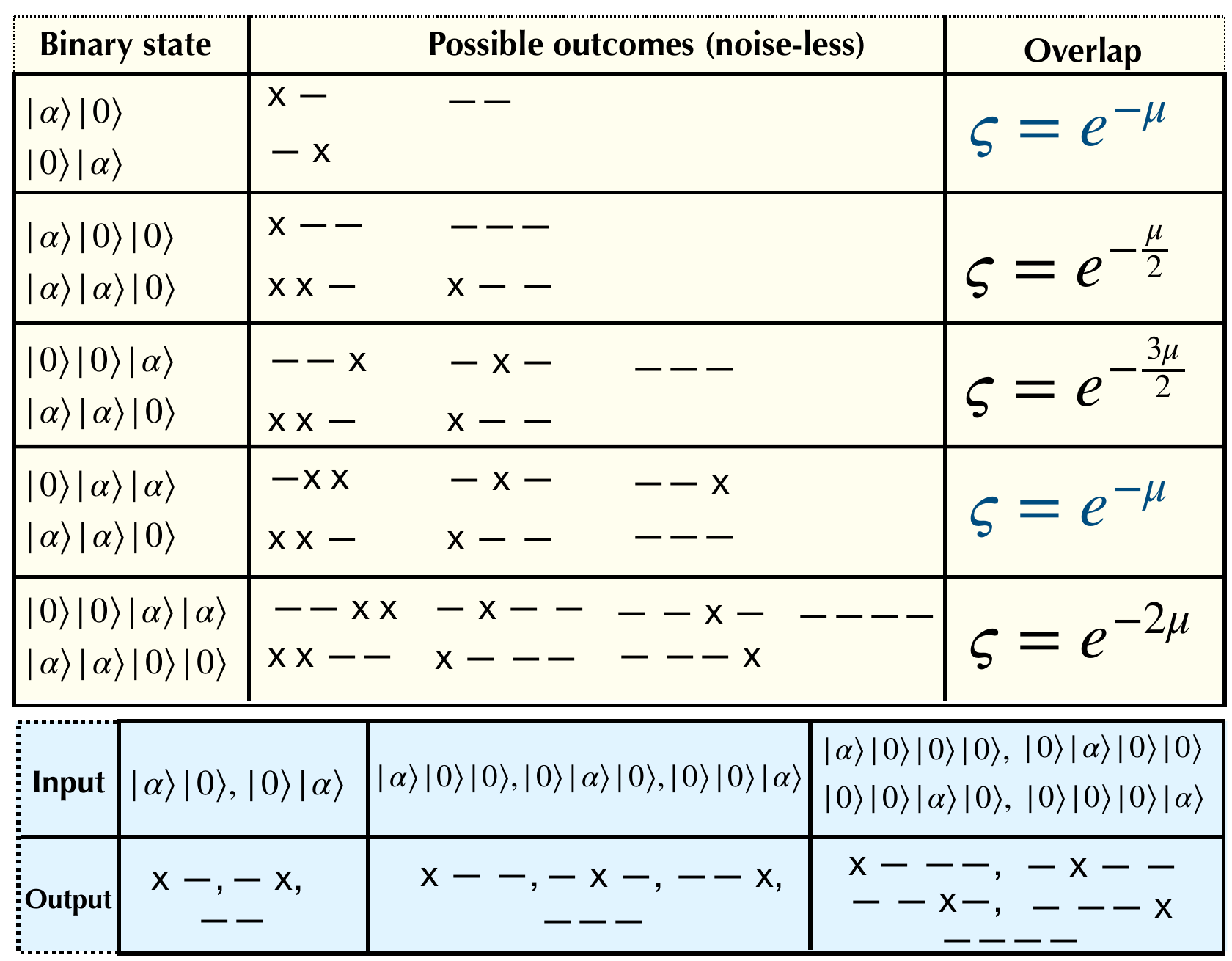}
    \caption{\textbf{Many-input vs Many-outcome approach.} 
    \textit{Top:} Many-outcomes approach with binary input; Examples of many-outcome scenarios with two input states. Note that the overlap value differs in each case.
    \textit{Bottom:} Many-input approach with categorizing the outcomes. Note: x and -- represent detection and no-detection events, respectively. 
    }
    \label{Fig::dout}
\end{table}
The many-outcome approach is studied for the continuous variable (CV) case in Ref. \cite{d_out} where the focus is on heterodyne and homodyne detectors with binary input.
In the time-bin encoding scheme, we can control the number of outcomes by adjusting the number of time-bins or the number of WCS in each configuration. It should be noted that the overlap bound is not considered in this argument and should be added as criteria when solving the SDP. As an example with dual input, it is shown in Fig. (\ref{Fig::optii}) that conditional entropy rises when the number of outcomes increases. 

As shown in table (\ref{Fig::dout})-top, the overlap could be different from case to case; this causes the optimal value of conditional min-entropy to take place at different mean-photon numbers; the inset of Fig. (\ref{Fig::optii}) shows the optimal mean-photon number as a function of outcomes for different overlaps. Similarly, a many-input case can be introduced while keeping the outcome minimal. Table (\ref{Fig::dout})-bottom shows examples of the possible setting of the many-input approach.

\subsection{Conditional Probability}
Given the inputs and the outputs, one can compute the input-output correlation by employing the conditional probability $p(b|x)$, i.e., the probability of receiving outcome b given input x:
\beq
p(b|x) = \sum\limits_\varsigma  {p_\varsigma} 
\tr[\hat\rho_x \hat\Pi_b^\varsigma],
\eeq
where $\hat\rho_x$ are the prepared states, $\hat\Pi_b^\varsigma$ are the POVMs describing the measurement,$\varsigma$ the classical variable provided to the adversary which describes the classical correlations between the experimental devices and the adversary. 

The detector's dark count rate (DCR) and ambient light are usually considered constant (on average); as they are independent of the incident photon's energy. However, the likelihood of obtaining an afterpulse click is directly related to the system's repetition rate.
Some detection events may not be caused by a WCS but could be afterpulses of an earlier detection event—the higher the system's repetition rate, the higher the chance of an afterpulse in the subsequent time-bins. Consequently, it is critical to consider the afterpulsing effect for practical situations. 

The probability of registering a detection event in the $T^{th}$ bin is mainly subject to the presence of a WCS in that bin and afterpulsing due to detections in the earlier bins.
Assuming that afterpulsing only happens due to a detection event in the immediate bin before, the probability of detection in bin $T$ can be written as:
\begin{equation}
\begin{split}
    P_{\alpha}^{T}(1) & =1-e^{-\eta_{\text{det}}L|\alpha|^2}+\epsilon+P_{ap}P_{\alpha}^{T-1}(1) \\
    & = 1-e^{-\eta_{\text{det}}L|\alpha|^2}\\
    & +\epsilon+P_{ap}( 1-e^{-\eta_{\text{det}}L|\alpha|^2}+\epsilon+P_{ap}P_{\alpha}^{T-2}(1)) \\
    & \cdots \\
    & = \frac{1-e^{-\eta_{\text{det}}L|\alpha|^2}+\epsilon}{1-P_{ap}}.
\end{split}
\label{EQ::P_11}
\end{equation}
where $P_{\alpha}^{T}(1)$ is the probability of registering a detection when sending $|\alpha\rangle$, $\eta_{\text{det}}$ and $L$ are detector efficiency and source-measurement loss, $\epsilon$ is for devices' imperfections and classical noises, e.g., dark counts, background noise, etc., and $P_{ap}$ represents the afterpulse probability due to a detection event at one bin distance which is the intrinsic character of a single-photon avalanche diode (SPAD) that can be characterized experimentally. In Eq. (\ref{EQ::P_11}), we substituted $P_{\alpha}^{T-2}(1)$ with its value and formed a geometric series to find the result.

The rest of the probabilities can be expressed as
\beq
\begin{aligned}
& P_{\alpha}(0)=1-P_{\alpha}(1) \\
& P_{\varnothing}(1)=P_{ap}(\frac{1-e^{-\eta_{\text{det}}L|\alpha|^2}+\epsilon}{1-P_{ap}})+\epsilon \\
& P_{\varnothing}(0)=1-P_{\varnothing}(1),
\end{aligned}
\label{sing_pro}
\eeq
where $P_{\alpha}(1)$, $P_{\varnothing}(1)$, \big( $P_{\alpha}(0)$, $P_{\varnothing}(0)$\big) represent the probability of registering a click \big(no-click\big) event when states $\ket{\alpha}$ and $\ket{0}$ are transmitted. Given Eqs. (\ref{EQ::P_11}) and (\ref{sing_pro}), we can compute all the possible conditional probabilities for any input-output dimension.

\subsection{Randomness Generation Rate}
Besides security, the randomness generation rate is another key parameter of any QRNG. We previously discussed the security estimation for the general case with multiple input-output in the presence of classical side information and noise and how it scales up. Here, we consider the eventual generation rate in the time-bin protocol.

For a weak coherent pulse source with repetition rate $f$, the input-state generation, comprised of $n$ time-bins, scales down as $f/n$. However, the extractable randomness is determined by $H_{\min}$, Eq. (\ref{EQ::H_min}), and the number of states available in an equivalence group in a $n_{m,s}$-configuration. Hence, the rate can be written as,
\begin{equation}
    R = \frac{f}{n} \cdot |n_{m,s}| \cdot H_{\min}(n_{m,s}, \upsilon, \eta_{\text{det}}, \mu_{\text{optimal}}),
    \label{EQ::GenRate}
\end{equation}
where $|n_{m,s}|$ is the cardinality of the input-state set and $H_{\min}(n_{m,s}, \upsilon, \eta_{\text{det}}, \mu_{\text{optimal}})$ is the maximum extractable entropy from that set considering optimal $\mu$, all the sources of noise, and detector efficiency. As discussed in section (\ref{subsec::SecEst}), a general solution for $H_{\min}$ considering all the parameters is not feasible to present and this quantity needs to be calculated and optimized for each case.

It should be noted that we assume $f$ being below the detector's dead-time to avoid missing a signal. Additionally, the analysis considers all the possible inputs and outcomes. The investigation would become more straightforward in the case of the many-input or many-outcome approaches.
\begin{figure}[h]
\centering
\includegraphics[width=\linewidth]{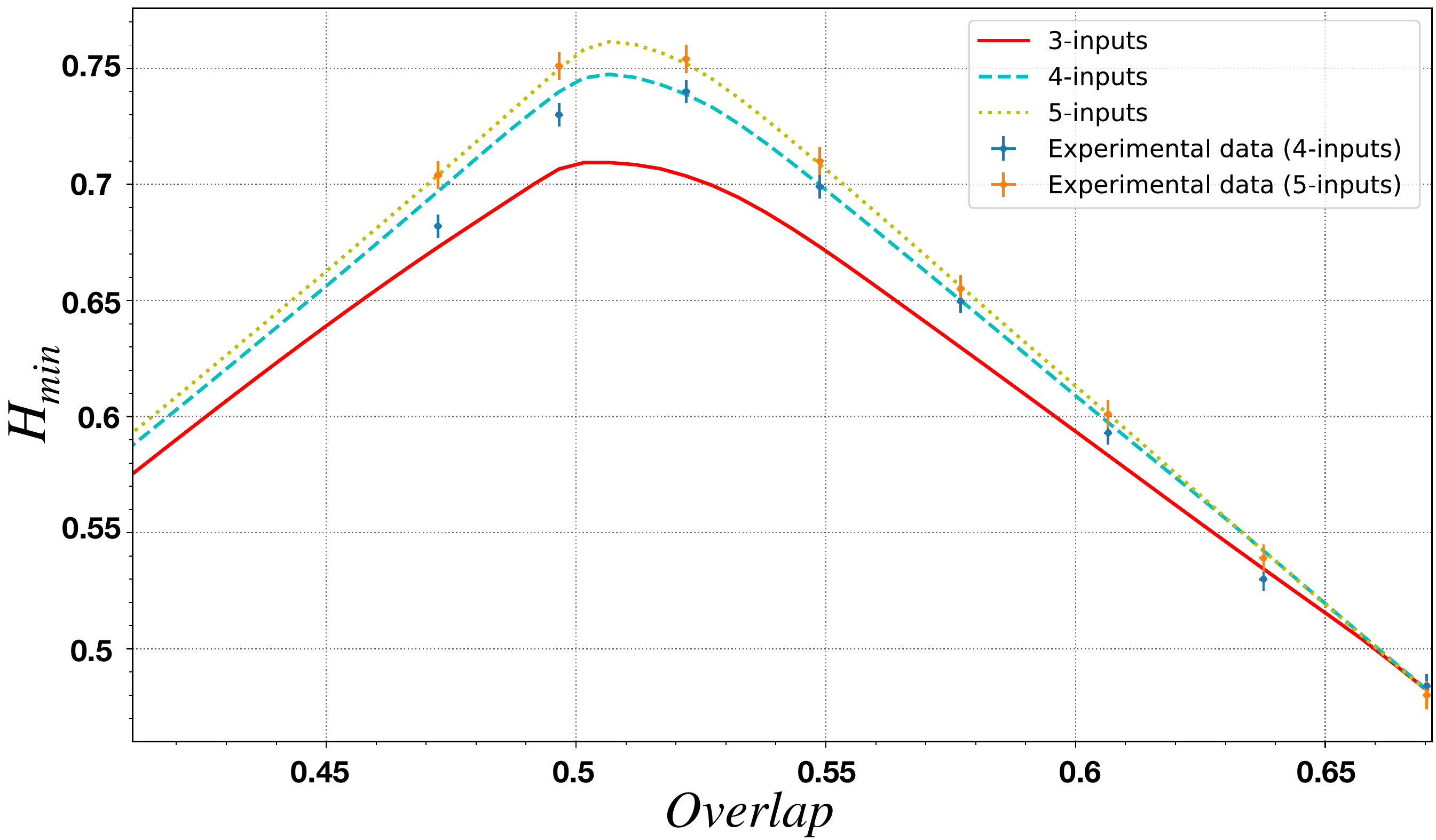}
\caption{Conditional min-entropy as a function of states' total overlap. The solid, dashed, and dotted curves represent the theoretical predictions for 3, 4, and 5 input configurations, respectively. At the same time, the blue and orange dots show the experimental data for 4 and 5 input cases measured with SPAD with 83 $\%$ efficiency.}
\label{Fig::res}
\end{figure}

\section{Experimental implementation}
This section investigates the experimental implementation and some practical considerations of this protocol. According to the protocol, the detection apparatus is considered a black box with no assumption on its performance. However, state generation must respect an overlap criteria, Eq. (\ref{EQ::OverlapCond}), which translates in two conditions; limited mean photon number $\mu$ per WCS and WCS positioning in an $n$-time-bin state.
\begin{figure}[h]
    \centering
    \includegraphics[width=\linewidth]{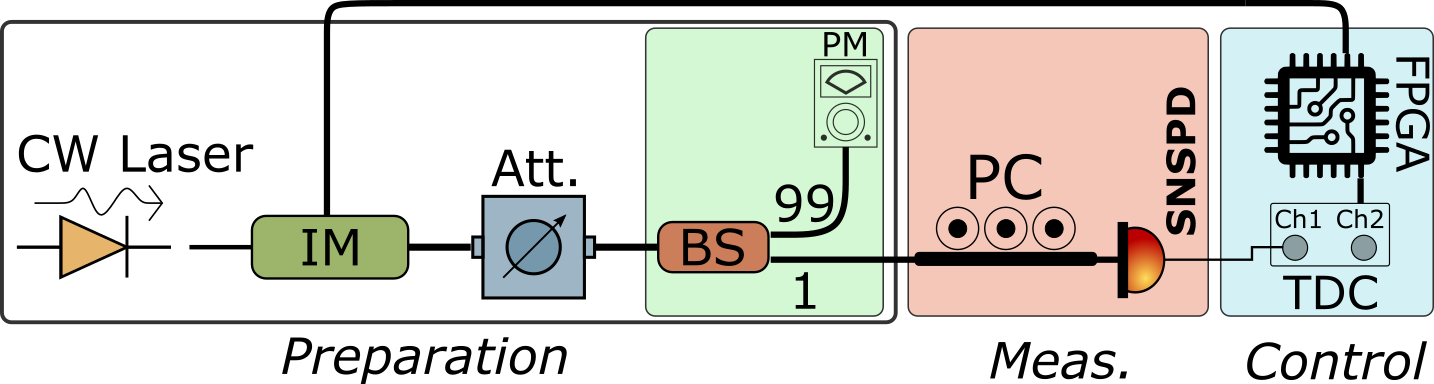}
    \caption{Schematic of the QRNG setup. A continuous wave laser (CW) is carved to form a train of pulses according to the protocol selected by the user. A combination of an attenuator (att.) and a 99:1 beamsplitter bring the power to single photon level where the 99\% output is monitored constantly with a power meter (PM) to certify the overlap condition. The single photons are then sent to a detector (SNSPD) for measurement. The polarization controller (PC) adjusts the polarization to maximize efficiency. The detection events are registered with a time-to-digital converter (TDC). State generation and measurement are governed and synchronized with the field programmable gate array (FPGA).}
    \label{Fig::chall}
\end{figure}

Fig. (\ref{Fig::chall}) shows a schematic representation of the setup. The $n$-time-bin state is generated by carving a 1550 nm continuous wave laser (CW) into pulses with 120 ps pulse width and a repetition rate of 31.25 MHz. Two cascaded intensity modulators, shown as one in the setup, guarantee high extinction ratio and perfect state generation. The repetition rate is chosen such that it matches the detector's dead-time and to minimize the chance of no-detection events. A field programmable gate array (FPGA) generates the electrical signal to drive the intensity modulators and to synchronize the measurement apparatus. To verify the overlap criteria, WCS placement is controlled such that the final state matches a subset, see Fig. (\ref{Fig::PermutStates}). A 99:1 beamsplitter separates the signal with the 99\% arm redirected to a power meter (PM). A variable optical attenuator (VOA) then sets the mean photon number to $\mu_{\text{optimal}}$ extracted from the security estimation process.

The quantum states are then sent and measured with a superconducting nanowire single photon detector (SNSPD) with ~30 ns dead-time, ~80 DCR, and ~83\% detection efficiency. The detection events are then registered with a time-to-digital converter (TDC) with 1 ps resolution and are analyzed for randomness extraction.

It is worth noting that in the time-bin encoding, detector's dead-time is the main limiting factor for high repetition rate state generation. 
\begin{figure*}[t!]
\centering
\includegraphics[width=\linewidth]{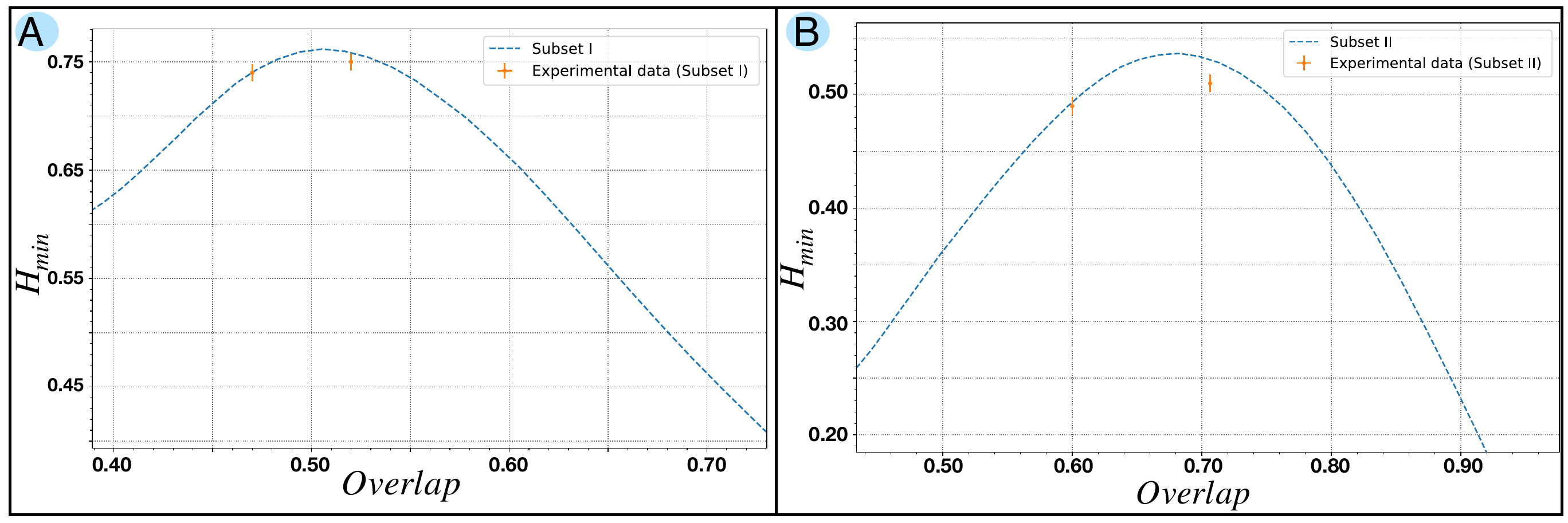}
\caption{Conditional min-entropy as a function of states' total overlap for Subsets I (A) and II (B) represented in Fig. \ref{Fig::PermutStates}. In both figures, the dashed line shows the theoretical predictions, and the orange dots represent the experimental data.}
\label{Fig::subsets}
\end{figure*}

\section{Results \& Discussion}
This section presents the theoretical and experimental min-entropy of different configurations, intending to validate the theoretical estimations. Foremost, the input-output correlation $P(b|x)$ is estimated by performing several measurements with various overlaps and gathering the detector's outcomes $b$ for given input $x$. The extractable amount of randomness is evaluated by inserting the input-output correlation and states' overlap into the SDP, which numerically computes the min-entropy.

We consider the simplest case: supplying one bin with a WCS and filling the rest of the bins with VS. Possible outcome configurations increase by raising the number of inputs, leading to a different input-output correlation and entropy. As shown in Fig. \ref{Fig::res}, the amount of extractable randomness conditioned on the classical side-information increases for the cases with a higher number of inputs.

Alternative forms of input configurations with more WCSs can also be considered. Paying attention to the 4-input case as an example, as shown in Fig. \ref{Fig::PermutStates}, instead of using the typical input configurations (set I, II, and III), one can implement subsets I and II, which require a ternary and binary initial seed rather than quartet one, downsizing the computational complexity, see Fig. \ref{Fig::com}. 
In Fig. (\ref{Fig::subsets}), the conditional min-entropy is plotted as a function of states' overlap for subsets I and II. The dashed curve is the expected theoretical results obtained for our experimental parameters which is in acceptable agreement with the experimental data taken from SNSPD with 83$\%$ detection efficiency and for various mean photon numbers.

The maximum conditional min-entropy for subsets I and II is $0.759$ and $0.546$, respectively, which are remarkably higher compared to typical binary and ternary input configurations at $\sim 0.2$ and $\sim 0.25$ obtained with detectors with 80\% and higher than 90\%  detection efficiencies, respectively \cite{Brask2017,Tebyanian_2021}. It should be noted that this higher rate entropy is achievable without the need of adjusting the optical setup and can be done in the signal preparation and post-processing stage. Furthermore, the randomness generation rate scaled from $0.11$ and $0.083$ to $0.1897$ and $0.136$ which is a considerable improvement achieved only by redefining the transmitted states.

\section{Conclusion}
In conclusion, we demonstrated a semi-DI QRNG based on the prepare-and-measure scenario exploiting a time-bin encoding scheme and single-photon detection technique investigating multiple input-output cases. Furthermore, the protocol is experimentally implemented using commercial-off-the-shelf components in a simple all-in-fibre optical setup at telecom wavelength, allowing a straightforward tunable input configuration needless of an optical switch.
We show that by holding the number of inputs(outcomes) fixed (minimal), known as the many-outcome (many-inputs) approach, one can increase the system entropy while keeping the computational complexity low. Additionally, a comprehensive study of time-bin encoding semi-DI QNRG is presented where, depending on the needs, one can select appropriate time-bin settings.

Besides, we compared this protocol's results with binary and ternary-input systems and showed that our protocol is capable of generating more randomness with the same optical setup.
The proposed protocol features advanced security since it only demands bounding the prepared states' overlap; the rest of the setup is not required to be characterized and can be classically correlated with the adversary. Alternatively, this protocol can be implemented in a different wavelength where single photon avalanche diodes (SPADs) have better detection efficiency, thus making this proposal chip-integrable.
In a nutshell, the semi-DI protocols' main advantage is to ease up the implementation complexity and enhance the generation rate preserving a high level of security. This paper demonstrates a semi-DI QRNG based on the overlap bound with an easy-to-implement experimental setup which can produce random numbers at a high rate with robust security applicable for various input-output configurations. 

\vspace{10pt}
\textbf{Acknowledgment:} This work is supported by the Center of Excellence SPOC - Silicon Photonics for Optical Communications (ref DNRF123), by the EraNET Cofund Initiatives QuantERA within the European Union’s Horizon 2020 research and innovation program grant agreement No. 731473 (project SQUARE), and by VILLUM FONDEN, QUANPIC (ref. 00025298). H. T. acknowledges the Innovate UK Industrial Strategy Challenge Fund (ISCF), project 106374-49229 AQuRand (Assurance of Quantum Random Number Generators).

\vspace{10pt}
\textbf{Conflict of interest statement:} The authors declare no conflicts of interest regarding this article.

\vspace{10pt}
\textbf{Data availability:}
The data that support the findings of this study are available from the corresponding author upon reasonable request.

\bibliography{bibliography}
\end{document}